# Evaluating Cache Coherent Shared Virtual Memory for Heterogeneous Multicore Chips

BLAKE A. HECHTMAN, Duke University
DANIEL J. SORIN, Duke University

The trend in industry is towards heterogeneous multicore processors (HMCs), including chips with CPUs and massively-threaded throughput-oriented processors (MTTOPs) such as GPUs. Although current homogeneous chips tightly couple the cores with cache-coherent shared virtual memory (CCSVM), this is not the communication paradigm used by any current HMC. In this paper, we present a CCSVM design for a CPU/MTTOP chip, as well as an extension of the pthreads programming model, called xthreads, for programming this HMC. Our goal is to evaluate the potential performance benefits of tightly coupling heterogeneous cores with CCSVM.

## 1 INTRODUCTION

The trend in general-purpose chips is for them to consist of multiple cores of various types—including traditional, general-purpose compute cores (CPU cores), graphics cores (GPU cores), digital signal processing cores (DSPs), cryptography engines, etc.—connected to each other and to a memory system. Already, general-purpose chips from major manufacturers include CPU and GPU cores, including Intel's Sandy Bridge [42][16], AMD's Fusion [4], and Nvidia Research's Echelon [18]. IBM's PowerEN chip [3] includes CPU cores and four special-purpose cores, including accelerators for cryptographic processing and XML processing.

In Section 2, we compare current heterogeneous multicores to current homogeneous multicores, and we focus on how the cores communicate with each other. Perhaps surprisingly, the communication paradigms in emerging heterogeneous multicores (HMCs) differ from the established, dominant communication paradigm for homogeneous multicores. The vast majority of homogeneous multicores provide tight coupling between cores, with all cores communicating and synchronizing via *cache-coherent shared virtual memory* (CCSVM). Despite the benefits of tight coupling, current HMCs are loosely coupled and do not support CCSVM, although some HMCs support aspects of CCSVM.

In Section 3, we develop a tightly coupled CCSVM architecture and microarchitecture for an HMC consisting of CPU cores and massively-threaded throughput-oriented processor (MTTOP) cores.[1] The most prevalent examples of MTTOPs are GPUs, but MTTOPs also include Intel's Many Integrated Core (MIC) architecture [31] and academic accelerator designs such as Rigel [19] and vector-thread architectures [21]. The key features that distinguish MTTOPs from CPU multicores are: a very large number of cores, relatively simple core pipelines, hardware support for a large number of threads per core, and support for efficient data-parallel execution using the SIMT execution model.

*We do not claim to invent CCSVM for HMCs; rather our goal is to evaluate one strawman design in this space.* As a limit study of CCSVM for HMCs, we prefer an extremely tightly coupled design instead of trying to more closely model today's HMC chips. We discuss many of the issues that arise when designing CCSVM for HMCs, including TLB misses at MTTOP cores and maintaining TLB coherence.

In Section 4, we present a programming model that we have developed for utilizing CCSVM on an HMC. The programming model, called *xthreads*, is a natural extension of pthreads. In the xthreads programming model, a process running on a CPU can spawn a set of threads on MTTOP cores in a way that is similar to how one can spawn threads on CPU cores using pthreads. We have implemented the xthreads compilation toolchain to automatically convert xthreads source code into executable code for the CPUs and MTTOPs.

In Section 5, we present an experimental evaluation of our HMC design and the performance of xthreads software running on it. The evaluation compares our full-system simulation of an HMC with CCSVM to a high-end HMC currently on the market from AMD. We show that the CCSVM HMC can vastly outperform the AMD chip when offloading small tasks from the CPUs to the MTTOPs.

---

[1] We use the term "GPU core" to refer to a streaming multiprocessor (SM) in NVIDIA terminology or a compute unit in AMD terminology.

In Section 6, we discuss the open challenges in supporting CCSVM on future HMCs. We have demonstrated the potential of CCSVM to improve performance and efficiency, but there are still issues to resolve, including scalability and maintaining performance on graphics workloads.

In Section 7, we discuss related work, including several recent HMC designs.

In this paper, we make the following contributions:

- We describe the architecture and microarchitecture for an HMC with CCSVM,
- We explain the differences between an HMC with CCSVM and state-of-the-art systems,
- We experimentally demonstrate the potential of an HMC with CCSVM to increase performance and reduce the number of off-chip DRAM accesses, compared to a state-of-the-art HMC running OpenCL, and
- We show how CCSVM/xthreads enables the use of pointer-based data structures in software that runs on CPU/MTTOP chips, thus extending MTTOP applications from primarily numerical code to include pointer-chasing code.

## 2 Communication Paradigms for Current Multicore Chips

In this section, we compare the tightly-coupled designs of current homogeneous multicores to the loosely-coupled designs of current heterogeneous multicores. At the end of this section, we focus on one representative heterogeneous multicore, AMD's Llano Fusion APU [4][2]; we use the APU as the system we experimentally compare against in Section 5. We defer a discussion of other HMCs until Section 7.

### 2.1 Homogeneous Multicore Chips

The vast majority of today's homogeneous chips [32][23][34][7], including homogeneous chips with vector units [31], tightly couple the cores with hardware-implemented, fine-grained, cache-coherent shared virtual memory (CCSVM). The virtual memory is managed by the operating system's kernel and easily shared between threads. Hardware cache coherence provides automatic data movement between cores and removes this burden from the programmer. Although there are some concerns about coherence's scalability, recent work shows that coherence should scale to at least hundreds of cores [28].

By coupling the cores together very tightly, CCSVM offers many attractive features. Tightly-coupled multicore chips have relatively low overheads for communication and synchronization. CCSVM enables software to be highly portable in both performance and functionality. It is also easy to launch threads on homogeneous CPUs since the kernel manages threads and schedules them intelligently.

### 2.2 Communication Options for Heterogeneous Multicore Chips

The most important design decision in a multicore chip is choosing how the cores should communicate. Even though the dominant communication paradigm in today's homogeneous chips is CCSVM, no existing HMC uses CCSVM. AMD, ARM, and Qualcomm have collaborated to create an architecture called HSA that provides shared virtual memory and a memory consistency model, yet HSA does not provide coherence [39]. Some chips, such as the Cell processor [17] and a recent ARM GPU[3], provide shared virtual memory but without hardware cache coherence. AMD also suggests that future GPUs may have hardware support for address translation [1]. Other HMC designs provide software-implemented coherent shared virtual memory at the programming language level [20][13]. Another common design is for the cores to communicate via DMA, thus communicating large quantities of data (i.e., coarse-grain communication) from one memory to another. These memories may or may not be part of the same virtual address space, depending on the chip design. The DMA transfer may or may not maintain cache coherence.

---

[2] We would have also considered Intel's SandyBridge CPU/GPU chip [42], but there is insufficient publicly available information on it for us to confidently describe its design.
[3] http://blogs.arm.com/multimedia/534-memory-management-on-embedded-graphics-processors/

The options for synchronizing between CPU cores and non-CPU cores are closely related to the communication mechanisms. Communication via DMA often leads to synchronization via interrupts and/or polling through memory-mapped I/O. Communication via CCSVM facilitates synchronization via atomic operations (e.g., fetch-and-op) and memory barriers.

### 2.3 A Typical HMC: The AMD Fusion APU

AMD's current Fusion APU, code-named Llano [11], is a heterogeneous multicore processor consisting of x86-64 CPU cores and a Radeon GPU. The CPU and GPU cores have different virtual address spaces, but they can communicate via regions of physical memory that are pinned in known locations. The chip has a unified Northbridge that, under some circumstances, facilitates coherent communication across these virtual address spaces. Llano supports multiple communication paradigms. First, Llano permits the CPU cores and the GPU to perform coherent DMA between their virtual address spaces. In a typical OpenCL program, the CPU cores use DMA to transfer the input data to the GPU, and the GPU's driver uses DMA to transfer the output data back to the CPU cores. Although the DMA transfers are coherent with respect to the CPU's caches, the load/store accesses by CPU cores and GPU cores between the DMA accesses are not coherent. Second, Llano allows a CPU core to perform high-bandwidth, uncacheable writes directly into part of the GPUs' virtual address space that is pinned in physical memory at boot. Third, Llano introduces the Fusion Control Link (FCL) that provides coherent communication over the Unified NorthBridge (UNB) but at a lower bandwidth. With FCL, the GPU driver can create a shared memory region in pinned physical memory that is mapped in both the CPU and GPU virtual address spaces. Assuming the GPU does not cache this memory space, then writes by the CPU cores and the GPU cores over the FCL are visible to each other, including GPU writes being visible to the CPUs' caches. A GPU read over the FCL obtains coherent data that can reside in a CPU core's cache. The FCL communication mechanism is somewhat similar to CCSVM, except (a) the virtual address space is only shared for small amounts of pinned physical memory and (b) the communication is not guaranteed to be coherent if the GPU cores cache the shared memory region.

## 3 CCSVM Chip Architecture and Microarchitecture

In this section, we present the clean-slate architecture and microarchitecture of a HMC with cache-coherent shared virtual memory. Where possible, we strive to separate the architecture from the microarchitecture, and we will highlight this point throughout the section. Throughout the rest of the paper, we assume that all cores are either CPU cores or MTTOP cores.

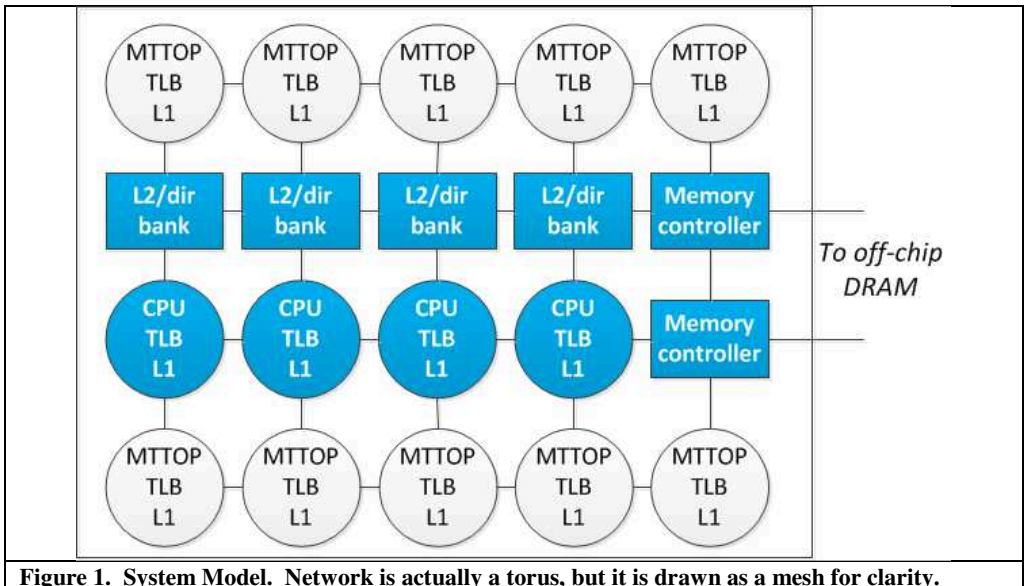

**Figure 1. System Model. Network is actually a torus, but it is drawn as a mesh for clarity.**

*Our HMC's CCSVM design is intentionally unoptimized and not tuned to the specific system model.* Where possible, we make conservative assumptions (e.g., our cache coherence protocol does not treat MTTOP cores differently from CPU cores, despite their known behavioral differences). Our goal is to isolate the impact of CCSVM without muddying the picture with optimizations. If unoptimized CCSVM outshines existing designs, then the difference is due to CCSVM itself and not due to any particular optimization.

### 3.1 Chip Organization

The chip consists of CPU cores and MTTOP cores that are connected together via some interconnection network. Each CPU core and each MTTOP core has its own private cache (or cache hierarchy) and its own private TLB and page table walker. All cores share one or more levels of globally shared cache. This cache is logically shared and CPU and MTTOP cores can communicate via loads and stores to this cache. This shared cache is significantly different than the *physically shared but logically partitioned* last-level cache in Intel's SandyBridge and IvyBridge chips; in the Intel chips, communication between CPU and GPU cores must still occur via off-chip DRAM.[4] We do not differentially manage the cache for fairness or performance depending on the core initiating the request [24].

We illustrate the organization of our specific microarchitecture in Figure 1, in which the CPU and MTTOP cores communicate over a 2D torus interconnection network (drawn as a mesh, rather than torus, for clarity). In this design, the shared L2 cache is banked and co-located with a banked directory that holds state used for cache coherence.

One detail not shown in the figure is our introduction of a simple controller called the MTTOP InterFace Device (MIFD). The MIFD's purpose is to abstract away the details of the MTTOP (including how many MTTOP cores are on the chip) by providing a general interface to the collection of MTTOP cores. The MIFD is similar to the microcontrollers used to schedule tasks on current MTTOPs. When a CPU core launches a task (a set of threads) on the MTTOP, it communicates this task to the MIFD via a write syscall, and the MIFD finds a set of available MTTOP thread contexts that can run the assigned task. Task assignment is done in a simple round-robin manner until there are no MTTOP thread contexts remaining. The MIFD does not guarantee that a task that requires global synchronization will be entirely scheduled, but it will write an error register if there are not enough MTTOP thread contexts available. The MIFD thus enables an architecture in which the number of MTTOP cores is a microarchitectural feature. The MIFD driver is a very simple piece of code (~30 lines), unlike drivers for current MTTOPs that perform JIT compilation from the HLL to the MTTOP's native machine language. The primary purposes of this driver are to assign threads to MTTOP cores, arbitrate between CPU processes seeking to launch MTTOP threads, and set up the virtual address space on the MTTOP cores.

### 3.2 Cache-Coherent Shared Virtual Memory

The key aspect of our architecture is to extend CCSVM from homogeneous to heterogeneous chips.

#### 3.2.1 *Shared Virtual Memory (SVM)*

<u>Architecture.</u> In all SVM architectures (for homogeneous or heterogeneous chips), all threads from a given process share the same virtual address space, and they communicate via loads and stores to this address space. The system translates virtual addresses to physical addresses, with the common case being that a load/store hits in the core's TLB and quickly finds the translation it needs. We assume that the caches are physically addressed (or at least physically tagged), as in all current commercial chips of which we are aware.

There are several possible differences between SVM architectures. For one, how are TLB misses handled? Some architectures specify that TLB misses are handled by trapping into the OS, whereas others specify that a hardware page table walker will handle the miss. Second, there is often architectural support for managing the page table itself, such as x86's CR3 register that is an architecturally-visible register that points to the root of the process's page table. Third, there is a range of architectural support for maintaining TLB coherence. Some architectures provide special instructions or interrupts to accelerate the process of TLB shootdown [38]. Because the CPU cores in the target system we study are x86, our HMC faithfully adheres to x86-specific architectural decisions, including the use of a hardware TLB miss handler (page table walker).

---

[4] http://www.anandtech.com/print/3922
http://www.realworldtech.com/page.cfm?ArticleID=RWT080811195102&p=9

Microarchitecture. Satisfying the x86 architecture, as it pertains to SVM, introduces some interesting issues in our heterogeneous chip. Adding TLBs and page table walkers to each MTTOP core is straightforward; we discuss the costs of adding them in Section 3.4. Sending the contents of the CR3 register to a MTTOP core when a task begins (as part of the task descriptor that is communicated via the write syscall to the MIFD) is also straightforward, although it does require us to add a CR3 register to each MTTOP core. However, two other issues are more complicated.

First, what happens when a MTTOP core encounters a page fault? The MTTOP core misses in its TLB and its page table walker identifies the miss as a page fault. On a CPU core, a page fault traps to the OS, but today's MTTOP cores are not (yet) running the OS. A relatively simple solution to this problem is to have the MTTOP core interrupt a CPU core, with the interrupt signaling that the CPU core should handle the page fault. We implement this mechanism via the MIFD, thus maintaining the abstraction that CPU cores do not need to be aware of specific MTTOP cores or of the number of MTTOP cores. The MIFD, as a device, may interrupt a CPU core on behalf of one of the MTTOP cores. One challenge, previously identified by IBM's PowerEN [12] and Intel's research prototypes [40][41], is that, unlike when a CPU core page faults, the interrupt occurs on the CPU core when the CPU core is not necessarily running the process that page faulted. The PowerEN skirts this problem by requiring the software to implement a user-level page fault handler on the CPU core. For CCSVM, we have the MTTOP core interrupt the CPU core with the interrupt cause (page fault) and the x86 CR3 register that is required to identify the necessary page table. AMD's IOMMU design [1] takes a somewhat similar approach of offloading GPU page faults to the CPU, but the request and acknowledgment are performed over PCIe.

Second, how do we keep the TLBs at the MTTOP cores coherent? TLB coherence in all-CPU chips is usually maintained via TLB shootdowns. A CPU core that modifies a translation sends an interrupt to other cores that may be caching that translation in their TLBs, and those TLBs invalidate their copies of the translation. On our chip, we must consider what happens when a CPU core initiates a shootdown. (A MTTOP core may not trigger a shootdown, since it does not run the OS and cannot modify translations.) We extend shootdown by having the CPU core signal the TLBs at all MTTOP cores to flush. Flushing is conservative, in that we could have only selectively invalidated entries, but it is a simple, viable option. A future consideration is to incorporate the TLBs into the coherence protocol [30].

### 3.2.2 Cache Coherence

Architecture. Cache coherence is not, strictly speaking, an architectural issue. To be precise, an architecture specifies only the memory consistency model (Section 3.2.3). However, virtually all homogeneous systems with SVM provide hardware cache coherence as a key element in supporting the consistency model.[5] Cache coherence protocols ensure that cores cannot continue to read stale values from their caches; protocols commonly enforce the "single writer or multiple readers" (SWMR) invariant [35].

Microarchitecture. Because coherence protocols are not architecturally visible, we can choose any protocol that is sufficient to support the desired consistency model. For our purposes, any protocol that maintains the SWMR invariant suffices. In our implementation, we choose a standard, unoptimized MOESI [37] directory protocol in which the directory state is embedded in the L2 blocks, similar to recent Intel and AMD chips [34][6]. With an inclusive L2 cache (as in Nehalem [34]), an L2 miss indicates that the block is not cached in any L1 and thus triggers an access to off-chip memory. Our choice of protocol is noteworthy only insofar as the protocol itself is not special. The assumption of write-back caches with coherence is a significant deviation from current MTTOP cores that generally have write-through caches with the ability to bypass the L1 for coherent reads, writes, and atomic operations. However, some current MTTOPs, such as MIC and ARM's Mali GPU [36], already provide cache-coherent shared memory or shared memory without cache coherence, like AMD's Heterogeneous System Architecture [39]. Recent academic research has also explored cache coherence tailored specifically for GPUs (without associated CPUs) [33]. We believe there is potential—in terms of performance and power—in tailoring a coherence protocol for a specific HMC, to take advantage of its distinctive features; we leave this research to future work.

---

[5] We do not consider software coherence [25] in this paper.

3.2.3 *Memory Consistency Model*

Homogeneous chips with CCSVM support a memory consistency model [35]. The consistency model specifies, at the architectural level, the legal orderings of loads and stores performed by threads in a process. For example, sequential consistency (SC) [22] specifies that all loads and stores must appear to perform in a total order, such that the total order respects the program order at each thread and each load obtains the value of the most recent store (in the total order) to the same address.

To the best of our knowledge, no heterogeneous chip has specified a clear memory consistency model (presumably because they do not support CCSVM), although high-level languages (HLLs) like CUDA and OpenCL provide ordering guarantees when programmers insert memory barrier operations (e.g., CUDA's thread fence and OpenCL's read and write fences). Now that we are considering CCSVM for heterogeneous chips, we believe it is time to specify an architectural-level consistency model for these chips. An architectural consistency model would help microarchitects to design systems by providing them with a clear specification of correctness. An architectural consistency model would also help programmers, either those who write in xthreads or those who write the software that supports HLLs.

The space of possible consistency models for arbitrary heterogeneous chips is enormous, because of the rich variety of memory access instructions and memory access patterns. We leave an exploration of this space for future work and instead, for now, conservatively provide a sequentially consistent (SC) model.[6] SC is the simplest model to reason about, and for the proof-of-concept design in this paper we are willing to forego many optimizations that violate SC (e.g., we have no write buffers between the cores and their caches). SC is far more restrictive than the memory models provided by current MTTOPs, particularly GPUs, which tend to prefer relaxed models for graphics computations. Recent work explores consistency models for homogeneous MTTOPs [15], but it is not clear how those results apply to HMCs that include MTTOPs. We are not arguing for SC, and we believe more relaxed models are likely preferable in the future, but we *are* advocating for the specification of precise consistency models.

3.2.4 *Synchronization*

Architecture: Our CCSVM architecture provides several synchronization primitives that can be used to construct typical HLL synchronization constructs (e.g., locks, barriers, signal/wait). In addition to the synchronization operations supported by the x86 CPU cores, the MTTOP ISA provides simple atomic operations like those in OpenCL (e.g., atomic_cas, atomic_add, atomic_inc, atomic_dec).

Microarchitecture: Today's MTTOP cores tend to perform atomic instructions at the last-level cache/memory rather than at the L1 cache as is often the case for CPUs. Because our goal is to supportt general-purpose MTTOP code, rather than just graphics code, and because we want the MTTOP cores to mesh well with the CPU cores, our MTTOP performs atomic operations at the L1 after requesting exclusive coherence access to the block.

## 3.3 CCSVM's Potential Benefits

The primary benefit of CCSVM is communication performance. As we show experimentally in Section 5, CCSVM can vastly outperform hardware that requires the CPUs and MTTOPs to communicate via off-chip DRAM. This result is intuitive; on-chip communication has far lower latency and far greater bandwidth. CCSVM provides a safe (coherent) mechanism for transforming the vast majority of communication to be on-chip rather than off-chip.

---

[6] Our x86 CPUs must satisfy the x86 consistency model [29]. Because the x86 model is more relaxed than SC, SC is a valid implementation of the x86 model.

CCSVM's shift to on-chip communication has secondary potential benefits, including lower energy and power costs (on-chip communication is more power- and energy- efficient than off-chip) and greater effective off-chip memory bandwidth (because this bandwidth is not being used for communication than can now be done on-chip).

The CCSVM architecture also offers two qualitative benefits. First, it has the virtue of being similar to the CCSVM architectures that industry and academia are both already familiar with. All other things being equal, familiarity is good. The community has a long history of designing, validating, and programming (homogeneous) CCSVM machines, and this accumulated experience and wisdom is a benefit when working on a new system. Second, CCSVM is an architecture with a clean, clear specification. Being able to incorporate all cores, regardless of type, in the architectural memory consistency model facilitates memory system design and validation, as well as programming.

### 3.4 CCSVM's Costs

Our CCSVM architecture and the specific microarchitecture we implement have their costs, with respect to today's HMCs. First, we have added page table walkers to each MTTOP. These are small structures, compared to the size of a MTTOP, but they are not free. Nvidia and AMD GPUs already have or soon will have TLBs [9], so we do not consider those an added cost for CCSVM. Second, we have added the MIFD, which is a single small structure—similar to the microcontroller in today's GPUs—for the entire chip. Third, we have extended the on-chip interconnection network to accommodate the MTTOPs. For the 2D torus in our microarchitecture, this extension is simply a matter of adding nodes in the torus to which we attach the MTTOPs; we do not need to increase the bandwidth of any switches or links to accommodate the MTTOPs. Fourth, we have added a few bits per cache block for coherence state. Each of these four hardware additions has area costs and energy/power costs, but none seem prohibitive. Without laying out an entire chip, though, we cannot quantify these costs. One can also argue that the hardware and power dedicated to CCSVM's additional structures represents an opportunity cost, because this hardware and power could be used for other purposes. Once again, we do not believe this opportunity cost is great; even if it costs us the same as one MTTOP core (which is significantly more than we expect), this is likely a reasonable tradeoff given the vast performance and power benefits CCSVM achieves.

One potential qualitative cost of CCSVM is its complexity. CCSVM is indeed a more complicated design than keeping the cores loosely coupled and having them communicate via DMA. Fortunately, the community has a long history of producing (homogeneous) CCSVM chips, and we believe industry can extend from homogeneous to heterogeneous chips.

Another potential drawback of CCSVM is a concern about coherence's scalability. Some architects believe that cache coherence does not scale to large numbers of cores, in which case proposing it as an option for future heterogeneous chips would indeed be problematic. However, based on recent analyses [10][28], we believe coherence will scale well enough for at least hundreds of cores and likely more.

### 3.5 Compatibility with Graphics and with CUDA/OpenCL

The primary use of many MTTOPs—specifically, GPUs—is still graphics and we expect it to remain this way for the foreseeable future. Thus we cannot allow our design, which is tailored for general-purpose computation, to hinder the graphics capability of a chip. There is also an established code base in CUDA and OpenCL that we cannot ignore. Fortunately our CCSVM design is compatible with graphics and other legacy code. Legacy code can run using the existing drivers and bypass our simple CCSVM driver. The caches can be reconfigured for use as scratchpads [27], and we can disable the cache coherence hardware and page table walkers at the MTTOPs. The interconnection network used for CCSVM is overkill for legacy code that has little communication between MTTOPs, but it poses no problem for running legacy code.

Table 1. Synopsis of Basic API Functions

| Called By | Function | Description |
|---|---|---|
| CPU | create_mthread(void* fn, args* fnArgs, ThreadID firstThread, ThreadID lastThread) | Spawns a set of one or more MTTOP threads that each runs the specified function with the specified arguments. |
| | wait(ConditionVar* condition, ThreadID firstThread, ThreadID lastThread, unsigned int waitCondition) | CPU thread sets array of condition variables to WaitingOnMTTOP and waits until MTTOP threads change array elements to Ready. The waitCondition argument could specify, for example, that the CPU wait for malloc requests from the MTTOP threads. |
| | signal(ConditionVar* condition, ThreadID firstThread, ThreadID lastThread) | CPU thread sets array of condition variables to Ready so that MTTOP threads can stop waiting |
| | cpu_mttop_barrier(ThreadID firstThread, ThreadID lastThread, barrierArray* barrierArray, bool* sense) | CPU thread waits for all MTTOP threads to write to array of barrier locations, then CPU flips sense. |
| MTTOP | wait(ConditionVar* condition, ThreadID firstThread, ThreadID lastThread) | MTTOP threads set array of condition variables to WaitingOnCPU and waits until CPU changes these array elements to Ready |
| | signal(ConditionVar* condition, ThreadID firstThread, ThreadID lastThread) | MTTOP threads set array of condition to Ready so that CPU thread can stop waiting |
| | cpu_mttop_barrier(barrierArray* barrierArray, bool* sense) | MTTOP thread writes to its barrier array entry, then waits for sense to flip. |
| | mttop_malloc(int size) | Returns dynamically allocated memory (like malloc) |

## 4 Xthreads Programming Model

Given that we now have a CCSVM design for an HMC, we need to be able to write software that takes advantage of having CCSVM. We developed xthreads with the goal of providing a pthreads-like programming model that exploits CCSVM and is easy to use. We could not simply adopt the current versions of OpenCL and CUDA, because they do not exploit CCSVM; however, OpenCL or CUDA could be implemented to run on top of xthreads.

We do not claim that xthreads is optimal in any sense, nor do we wish to quantitatively compare it to other programming models for heterogeneous chips. Rather, we have developed xthreads as a proof-of-concept—that is, there exists at least one reasonable programming model that enables us to demonstrate the benefits of CCSVM. Future work will delve more deeply into improving and evaluating the programming model.

### 4.1 API

The xthreads API, summarized in Table 1, extends pthreads by enabling a thread on a CPU core to spawn threads on non-CPU cores. For our system model in which the non-CPU cores are MTTOP cores, the API's create_mthread ("m" is for "MTTOP") function enables a CPU thread to spawn a set of SIMT threads that will run on one or more MTTOP cores (depending on the microarchitecture). This function is roughly equivalent to (a) calling OpenCL's clEnEnqueueNDRange() with a work group size equal to the number of threads or (b) invoking a CUDA kernel with one thread block. Given that CCSVM provides coherence across all threads and all memory, there is no use having xthreads provide more levels of task splitting. Unlike CUDA and OpenCL, xthreads does not require that a subset of threads needs to be located on the same MTTOP core in order to enable communication between threads. Thus even small tasks can be split up to use the entire MTTOP instead of one MTTOP core.

The two primary mechanisms for synchronization are wait/signal and barrier. A CPU thread or a set of MTTOP threads can wait until signaled by CPU or MTTOP threads. The wait/signal pair operates on condition variables in memory. For example, a CPU thread can wait for an array of $C$ condition variables to all change from WaitingOnMTTOP to Ready, and this transition will occur when $C$ MTTOP threads have changed these condition variables to be Ready. The barrier function is a global barrier across one CPU thread and a set of MTTOP threads. We considered other types of barriers (e.g., a barrier that applies only to MTTOP threads) and a lock, but we found that we did not need them for our algorithms.

### 4.2 Compilation Toolchain

We have implemented a compilation framework that enables us to easily convert xthreads source code into an executable that runs on both the CPU and MTTOP cores. We illustrate the compiler toolchain in Figure 2. This toolchain and the executable model (i.e., embedding the MTTOP's code in the text segment of the CPU's executable) are quite similar to the CHI prototype [40].

### 4.3 Runtime

When a process begins, the CPU cores begin executing threads as they would in a "normal" pthreads application. The differences begin when a CPU thread calls create_mthread() for a MTTOP task with $N$ threads. For each task, the library performs a write syscall to the MTTOP interface device. This write syscall describes a task as: {program counter of function, arguments to function, first thread's ID, CR3 register}. The MIFD then assigns incoming tasks to available MTTOP cores. If a task specifies more threads than the SIMD width of a single MTTOP core, then the task is distributed in SIMD-width chunks to multiple MTTOP cores. A SIMD-width chunk is known as a warp in NVIDIA terminology and a wavefront in AMD terminology.

When a MTTOP core receives a task from the MIFD, it sets its CR3 register so that it can participate in virtual address translation with the other threads (CPU and MTTOP) in the process. The MTTOP core then begins executing from the program counter it receives. When it reaches an Exit instruction, it halts and waits for the MIFD to send a new task.

### 4.4 Example of Xthreads vs. OpenCL

To illustrate the xthreads programming model and API, we provide a simple example. In Figure 4 and Figure 3, we provide the xthreads and OpenCL code, respectively, for computing the sum of two vectors of integers. The xthreads code is fairly simple and intuitive. There is a minimum amount of overhead to create tasks and set up the MTTOP. The OpenCL code is far more complicated, largely because of all of the work it has to do to explicitly communicate between the CPU and the MTTOP. Increased code complexity obviously does not directly lead to poorer performance, but it does reveal situations in which more work must be done.

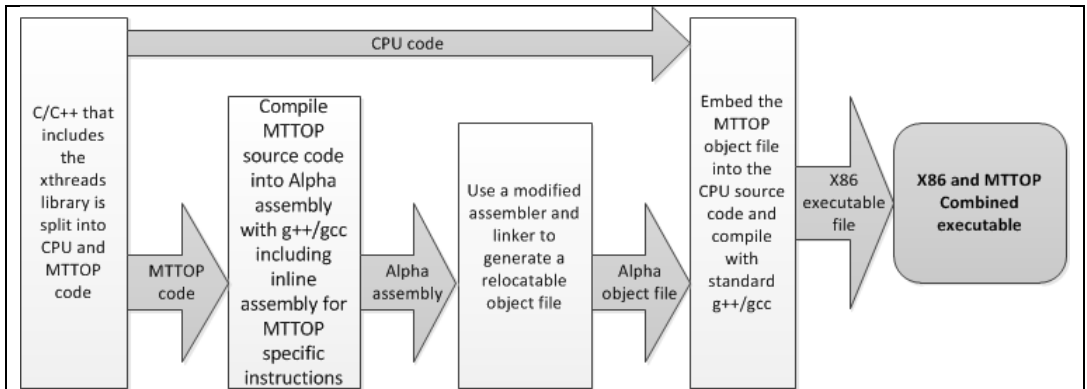

**Figure 2. Compilation Toolchain**

**In the kernel file**

```
__kernel void vector_add(__global __read_only int * v1,
                         __global __read_only int * v2,
                         __global __write_only int * sum){
    unsigned int tid = get_global_id(0);
    sum[tid] = v1[tid] + v2[tid];
}
```

**In the host file**

```
int main(){
  cl_platform_id platform_id = NULL;
  cl_device_id device_id = NULL;
  cl_uint ret_num_devices, ret_num_platforms;
  cl_int ret;
  ret = clGetPlatformIDs(1, &platform_id, &ret_num_platforms);
  ret = clGetDeviceIDs( platform_id,CL_DEVICE_TYPE_DEFAULT,1,&device_id, &ret_num_devices);
          // Create an OpenCL context
  cl_context context = clCreateContext( NULL, 1, &device_id, NULL, NULL, &ret);
          // Create a command queue
  cl_command_queue cmd_queue = clCreateCommandQueue(context, device_id, 0, &ret);
  cl_program program = clCreateProgramWithSource(context, 1, (const char **)&source_str, (const size_t *)&source_size, &ret);
          // Build the program
  ret = clBuildProgram(program, 0/*1*/, 0/*&device_id*/, NULL, NULL, NULL);
  cl_mem v1_mem_obj = clCreateBuffer(context,
      CL_MEM_ALLOC_HOST_PTR |
      CL_MEM_READ_WRITE,256*sizeof(int), NULL, &ret);
  cl_mem v2_mem_obj = clCreateBuffer(context,
      CL_MEM_ALLOC_HOST_PTR |
      CL_MEM_READ_WRITE, 256*sizeof(int), NULL, &ret);
  cl_mem sum_mem_obj = clCreateBuffer(context,
      CL_MEM_ALLOC_HOST_PTR |
      CL_MEM_READ_WRITE, 256*sizeof(int), NULL, &ret);
  int *v1 = (int*)clEnqueueMapBuffer(cmd_queue, v1_mem_obj,
      CL_TRUE, 0, 0, 256*sizeof(int),0, NULL, NULL,NULL);
  int *v2 = (int*)clEnqueueMapBuffer(cmd_queue, v2_mem_obj,
      CL_TRUE, 0, 0, 256*sizeof(int),0, NULL, NULL,NULL);
  for(int i = 0; i < 256; i++) {
    v1[i] = rand();
    v2[i] = rand();
  }
  clEnqueueUnmapMemObject(cmd_queue,v1_mem_obj,a,0,NULL,NULL);
  clEnqueueUnmapMemObject(cmd_queue,v2_mem_obj,b,0,NULL,NULL);
  cl_kernel kernel = clCreateKernel(program, "vector_add", &ret);
          // Execute the OpenCL kernel on the list
  size_t global_item_size = size ; // Process the entire lists
  size_t gsize = size ; // Process the entire list
  size_t local_item_size = (size<64)?size:64; // Process one item at a time
  global_item_size = global_item_size/local_item_size;
  cl_event x;
  ret=clSetKernelArg(kernel, 0, sizeof(int), &size);
  ret = clSetKernelArg(kernel, 1, sizeof(cl_mem), (void *)&v1_mem_obj);
  ret = clSetKernelArg(kernel, 2, sizeof(cl_mem), (void *)&v2_mem_obj);
  ret = clSetKernelArg(kernel, 3, sizeof(cl_mem), (void *)&sum_mem_obj);
  ret = clEnqueueNDRangeKernel(cmd_queue, kernel, 1, NULL, &gsize, NULL, 0, NULL,NULL);
  clFinish(cmd_queue);
  clEnqueueUnmapMemObject(cmd_queue,sum_mem_obj,sum,0,NULL,NULL);
  clReleaseMemObject(v1_mem_obj);
  clReleaseMemObject(v2_mem_obj);
  clReleaseMemObject(sum_mem_obj);
  return 0;
}
```

**Figure 3.  OpenCL Code**

```
struct args{
   int* v1, v2, sum;  // 3 vectors of ints
   bool* done;
}

_MTTOP_  void add(int tid, args* arg){
   arg->sum[tid]=arg->v1[tid] +arg->v2[tid];
   mthread_signal(arg->done);
}

_CPU_ int  main(int argc, char** argv){
   args inputs;
   inputs.v1=malloc(256*sizeof(int));
   inputs.v2=malloc(256*sizeof(int));
   inputs.sum=malloc(256*sizeof(int));
   inputs.done=malloc(256*sizeof(bool));
   for(int i=0;i<256;i++){
      inputs.v1[i]=rand();
      inputs.v2[i]=rand();
      inputs.done[i]=0;
   }
   mthread_create(0,256,&add,&inputs);
   mthread_wait(0,255,inputs.done);
   free(inputs.v1);
   free(inputs.v2);
   free(inputs.sum);
   free(inputs.done);
}
```

**Figure 4. Xthreads Code**

## 5 Experimental Evaluation

The goal of this evaluation is to quantitatively determine the viability of the CCSVM system model for heterogeneous chips, for at least the specific implementation of it we have presented, and the xthreads programming model. We believe this is the first experimental evaluation of the impact of CCSVM on performance, memory access efficiency, and programmability. We also believe that this is the first experimental evaluation of a CPU/MTTOP chip executing code with pointer-based data structures. Throughout this evaluation, we focus on the memory system and communication; we try to factor out the details of the CPU and MTTOP core pipelines, which are orthogonal to this work.

### 5.1 Methodology and Target System

We have implemented our chip design in the gem5 full-system simulator [2]. Our extensions to gem5 enable it to faithfully model the functionality and timing of the entire system. The simulated CPU cores are in-order x86 cores that run unmodified Linux 2.6 with the addition of our simple MIFD driver (~30 lines of C code). The MTTOP cores are SIMT cores that have an Alpha-like ISA that has been modified to be data parallel. The MTTOP's ISA is also similar to PTX, the assembly-like intermediate language to which CUDA is compiled and that Nvidia drivers convert to the native ISA of Nvidia GPUs. The details of the target system are in **Table 2**.

Evaluating the performance of CCSVM running xthreads benchmarks is somewhat challenging for two reasons. First, no xthreads benchmarks existed prior to this work, so we have had to port and re-write benchmarks. Second, comparing the performance of CCSVM to other system models is complicated by the different programming models and the complexity of modeling existing systems. For example, to truly model Intel's SandyBridge or an AMD Fusion chip in a comparable way, we would need access to its driver and its native ISA, among other information that is not publicly available. We fundamentally cannot simulate a currently available HMC and, as academics, we cannot implement a complete HMC in hardware.

**Table 2. Simulated CCSVM System and AMD System Configurations**

|  | **CCSVM System (simulated)** | **AMD APU (A8-3850 hardware) [11]** |
|---|---|---|
| CPU | 4 in-order x86 cores, 2.9 GHz, max IPC=0.5 | 4 out-of-order x86 cores, 2.9GHz, max IPC=4 |
| MTTOP | • 10 MTTOP cores with Alpha-like ISA, 600MHz.<br>• Combined max of 80 operations per cycle<br>• Each MTTOP core supports 128 threads and can simultaneously execute 8 threads. | • 5 SIMD processing units with 16 VLIW Radeon cores per SIMD unit, 600 MHz.<br>• Combined max of 80 VLIWinstrs/cycle<br>• Each VLIW instruction is 1-4 operations (max 320 operations per cycle). |
| On-chip memory | • Each CPU core has:<br>  L1I, L1D: write-back, 64KB, 4-way, 2-cycle hit<br>  TLB: 64-entry, fully-associative<br>• Each MTTOP core has:<br>  L1I, L1D: write-back, 16KB, 4-way, 1-cycle hit<br>  TLB: 64-entry, fully-associative<br>• All CPU and MTTOP cores share an inclusive 4MB L2:<br>  4 1MB banks, 10 CPU cycles, 2 MTTOP cycles | • Each CPU core has:<br>  L1I, L1D: 64KB, 4-way, 1ns hit<br>  L2: 1 MB, 3.6ns hit<br>  L2 TLB: 1024-entry<br>• Each SIMD processing unit has:<br>  32KB of local memory |
| Off-chip memory | 2GB DRAM, hit latency 100ns | 8GB DDR3 DRAM, hit latency 72ns |
| On-chip network | 2D torus, 12 GB/s link bandwidth | CPUs connected to each other via crossbar CPUs and GPUs fully connected to memory controllers |

Thus, we cannot perform a perfectly equivalent "apples-to-apples" comparison. Instead, we compare a real, current HMC to a simulator of our CCSVM design and configure the simulator to be conservative so as to favor the current HMC over our design. We purchased an AMD "Llano" system based on its Quad-Core A8-3850 APU [11], and we use this real hardware running OpenCL software for comparisons to CCSVM running xthreads software. The APU's specifications are in **Table 2**.

The configuration parameters show how we favor the APU over the CCSVM design. Notably, the simulated CCSVM CPU cores are in-order and capable of executing only one instruction every other cycle (i.e., max IPC is 0.5); thus, the simulated CCSVM CPU cores perform far worse than the CPU cores in the APU. Also, the maximum throughput of the simulated CCSVM MTTOP is less than that of the APU's GPU by a factor equal to the utilization of the VLIW width of the APU's GPU cores. When the APU's GPU is fully utilizing each VLIW instruction, it has a throughput that is 4x that of the simulated CCSVM's MTTOP; when the APU's GPU is at its minimum VLIW utilization, its throughput is equal to that of the CCSVM MTTOP.

### 5.2 Performance on "Typical" General Purpose Benchmarks

We have argued that the loose coupling between the CPU and MTTOP cores in today's systems is inefficient. To experimentally demonstrate the potential benefits of tighter coupling between the CPU and MTTOP cores, we first compare the execution of a (dense) matrix multiplication kernel that is launched from a CPU to as many MTTOP cores as can be utilized for the matrix size. Intuitively, the overhead to launch a task will be better amortized over larger task sizes, and the benefit of CCSVM will be highlighted by how it enables smaller tasks to be profitably offloaded to the MTTOP cores. In Figure 5, we plot the *log-scale* runtimes of the AMD APU running OpenCL code and CCSVM running xthreads code, relative to the AMD CPU core (i.e., just using the CPU core on the APU chip), as a function of the matrix sizes. For the APU, we present two runtime datapoints: full runtime and runtime without compilation and without OpenCL initialization code.

The results are striking: CCSVM/xthreads greatly outperforms the APU, especially for smaller matrix sizes. Eventually, as the matrices reach 1024x1024, the APU's performance catches up to CCSVM/xthreads, because the APU's raw GPU performance exceeds that of our simulated

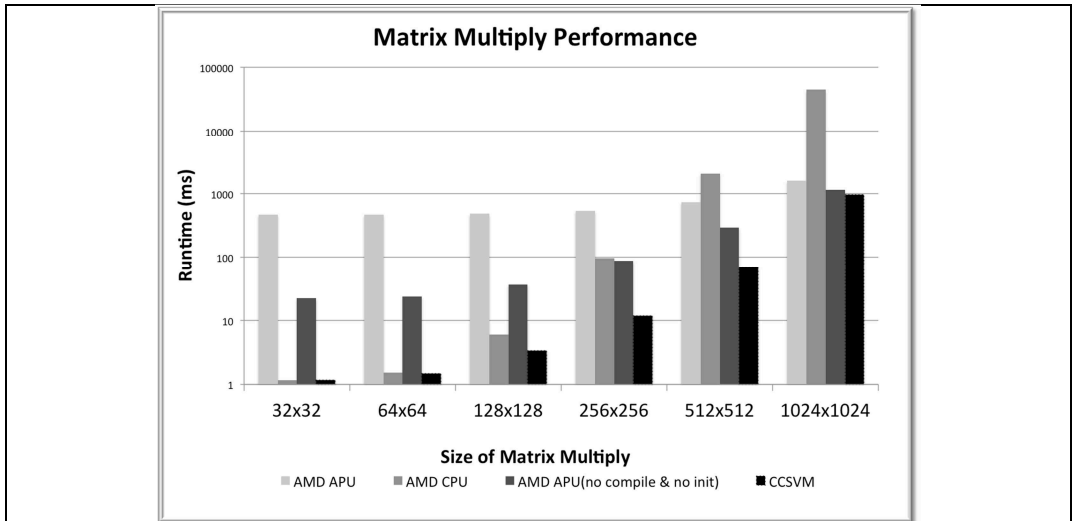

**Figure 5. Performance on Matrix Multiply. Results show how CCSVM reduces overhead to launch MTTOP tasks.**

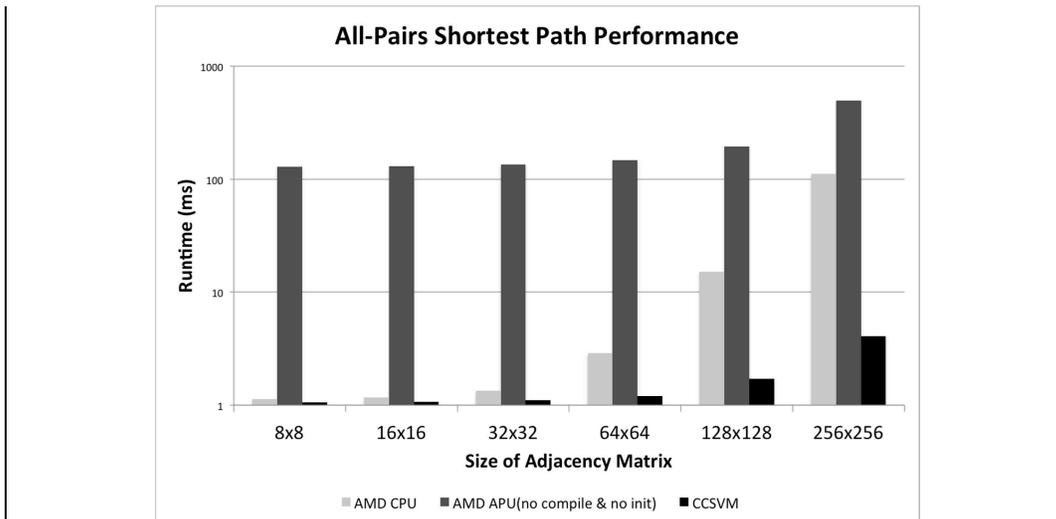

**Figure 6. Performance on All-Pairs Shortest Path. Results show how CCSVM improves performance by avoiding multiple MTTOP task launches for each parallel phase.**

MTTOP cores. The results dramatically confirm that optimizing the communication between the CPU and MTTOP cores offers opportunities for vastly greater performances and for profitably offloading smaller units of work to MTTOP cores, thus increasing their benefits.

In Figure 6, we plot the log-scale runtime results for a different benchmark: all-pairs shortest path. The algorithm is a triply-nested loop that fills out an adjacency matrix—each *(x,y)* entry of the adjacency matrix is the distance between nodes *x* and *y* in a directed graph—with the shortest path from each node to each other node. The algorithm requires a barrier between each iteration of the outermost loop. Because the APU's synchronization is quite slow, the APU's performance never exceeds that of simply using the CPU core. As with the matrix multiplication benchmark, CCSVM/xthreads vastly outperforms the APU/OpenCL over a range of matrix sizes. Even after factoring out the OpenCL compilation and initialization time, CCSVM outperforms the APU by approximately two orders of magnitude.

### 5.3 Performance on "Atypical" General Purpose Benchmarks

We now explore two benchmarks that highlight the ability of CCSVM/xthreads to extend the types of computations that can be performed on CPU/MTTOP chips.

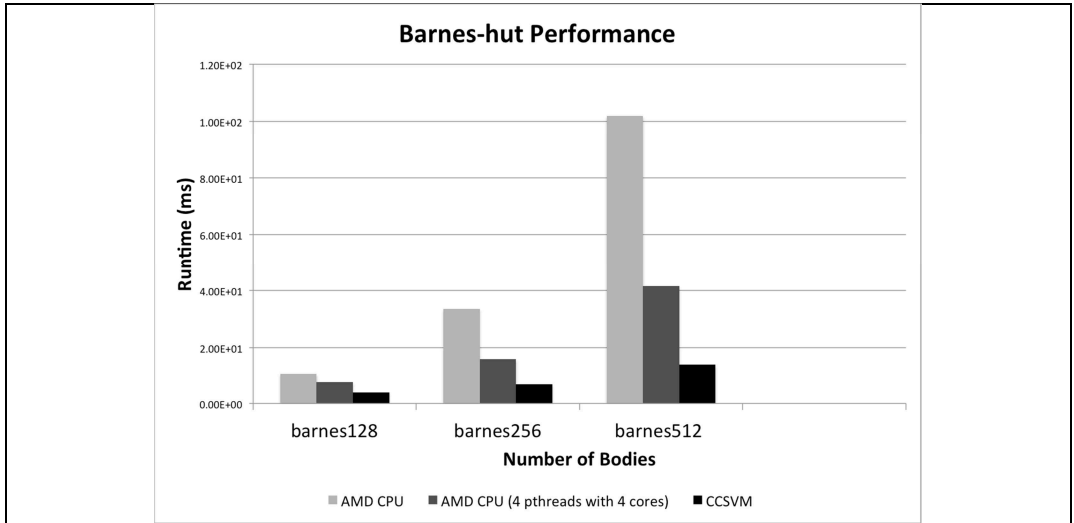

**Figure 7. Barnes-Hut performance. CCSVM/xthreads enables pointer chasing code.**

### 5.3.1 *Pointer-Based Data Structures, Recursion, and Frequent CPU-MTTOP Communication*

We ported the well-known barnes-hut n-body simulation benchmark from a pthreads version. This benchmark extensively uses pointers and recursion and, most problematically for current CPU/MTTOP chips, involves frequent toggling between sequential and parallel phases. Ideally, we would like to use the CPU core for the sequential phases and the MTTOP cores for the parallel phases, but switching between them on current HMCs is too slow to be viable. However, with CCSVM/xthreads, this switching between them and the associated CPU-MTTOP communication is fast and efficient. In Figure 7, we show the runtime of CCSVM/xthreads compared to a single AMD CPU core. We also compare to the pthreads version of the benchmark running with 4 threads on the 4 CPU cores on the AMD APU. We could not find or develop an OpenCL version of this benchmark against which to compare, so we could not exploit the APU's GPU. We are aware that there are techniques to perform the n-body problem without pointer-chasing [5], but we use the barnes-hut application as a "strawman" for parallelizable pointer-based applications. The results show that CCSVM/xthreads can outperform pthreads even for small problem sizes. There is great potential to use MTTOP cores to accelerate important algorithms, and this potential is unlocked by an HMC with CCSVM running xthreads code that exploits it.

### 5.3.2 *Pointer-Based, Dynamically Allocated Data Structures*

CCSVM with xthreads enables programmers to use space-efficient, pointer-based, dynamically allocated shared data structures. The only CPU-MTTOP software that partially provides this capability is CUDA 4.0 (released May 2011), which provides dynamic allocation for a subset of data types. Unless using CUDA 4.0, programmers convert data structures that would typically be pointer-based and dynamically allocated into statically allocated array-based structures, often at a huge expense in storage.

To support these types of benchmark, we added a mttop_malloc() function to the xthreads API. The mttop_malloc() function offloads the malloc to a CPU by having the CPU wait for the MTTOP threads to signal the CPU that they wish to dynamically allocate memory. When the CPU receives these signals, it performs the malloc() functions on their behalf and returns pointers to the MTTOP threads. This approach to dynamically allocating memory at the MTTOP is not particularly efficient or elegant, but it does demonstrate that it is possible; future work will involve optimizing this process.

To demonstrate the viability of dynamic allocation on CCSVM/xthreads, we developed a sparse matrix multiplication benchmark. For extremely large, sparse matrices, the only tractable way to represent them is with pointer-based data structures that link non-zero elements. In Figure 8, we plot the speedup of CCSVM/xthreads on this benchmark, relative to the performance on the AMD CPU. As with barnes-hut, there is no OpenCL version. The figure on the left has a fixed sparsity (1%) and varies the matrix size, and the figure on the right has a fixed matrix size and varies the sparsity. We observe that CCSVM/xthreads obtains speedups until the matrix density increases to the point at which the mttop_malloc() calls constrain the performance CCSVM/xthreads. Even where CCSVM has speedups, the speedups are not nearly as large as for the other benchmarks, due to the inefficiency of mttop_malloc() calls. However, our primary goal was not to demonstrate huge speedups, but rather to show that it is possible to run code with dynamically-allocated pointer-based data structures on a CPU/MTTOP chip.

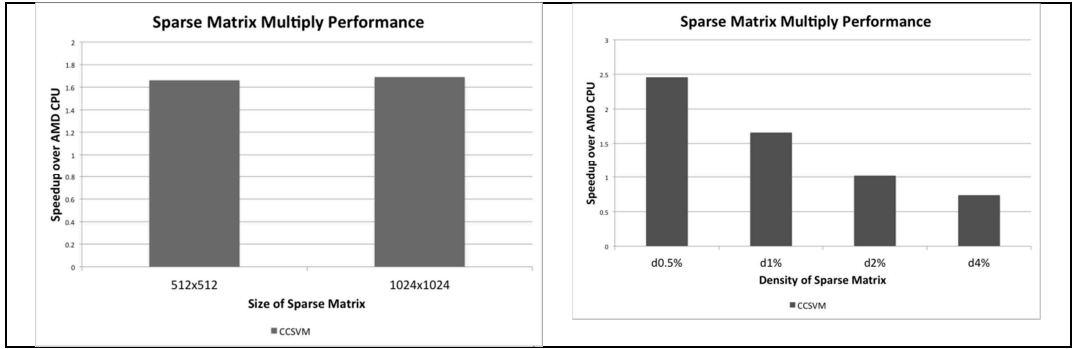

**Figure 8. Performance of Sparse Matrix Multiplication**

## 5.1 Off-Chip Bandwidth

CCVSM avoids the vast amount of off-chip traffic that current chips require for CPU-MTTOP communication. On a CCSVM chip, as in today's homogeneous all-CPU chips, the majority of communication is performed on-chip. In this section, we compare the number of off-chip accesses for our CCVSM chip and the APU, and we use the (dense) matrix multiply benchmark from earlier in this section. We obtain the results for the AMD system from its performance counters. We plot the *log-scale* results in Figure 9. As with the performance results, the differences between the APU/OpenCL and CCSVM/xthreads are dramatic. Furthermore, as the problem size increases, that ratio remains roughly the same, and the number of DRAM accesses from the AMD CPU core increases greatly as the working set outgrows the CPU core's caches. The CPU core, unlike the APU, cannot coalesce strided memory accesses, and thus its off-chip DRAM accesses increase far more than those of the APU.

These DRAM access results have two implications. First, the results help to explain the performance results. The APU requires far more DRAM accesses, and these long-latency off-chip accesses hurt its performance, relative to CCSVM and its largely on-chip communication. Second, given both the importance of using DRAM bandwidth efficiently and the energy consumed by DRAM accesses, the results show that CCSVM/xthreads offers tremendous advantages for system design.

## 6 Open Challenges in CCSVM for HMCs

The results in the previous section show that an HMC with a tightly-coupled CCSVM memory system can perform much better on small problem sizes compared to a state-of-the-art commercial HMC with looser coupling of cores. We now address some of the open challenges in achieving this performance benefit.

### 6.1 Maintaining Graphics Performance

Many current HMCs are CPU/GPU chips, and we must be careful to avoid harming the GPU performance on graphics. The coherence protocol for our HMC with CCSVM relies on a shared, inclusive L2 cache. It may be difficult to share such a low-level cache or even any level of cache, because current CPU caches are optimized for latency and current GPU caches are optimized for throughput. Given the importance of graphics performance, it seems unlikely that GPU caches will be modified to become more like CPU caches. Without sharing a cache, we would need an efficient technique to move data between the CPU and GPU cache hierarchies without accessing off-chip memory. A directory protocol with a stand-alone directory structure (i.e., not embedded in any cache) is a potential solution.

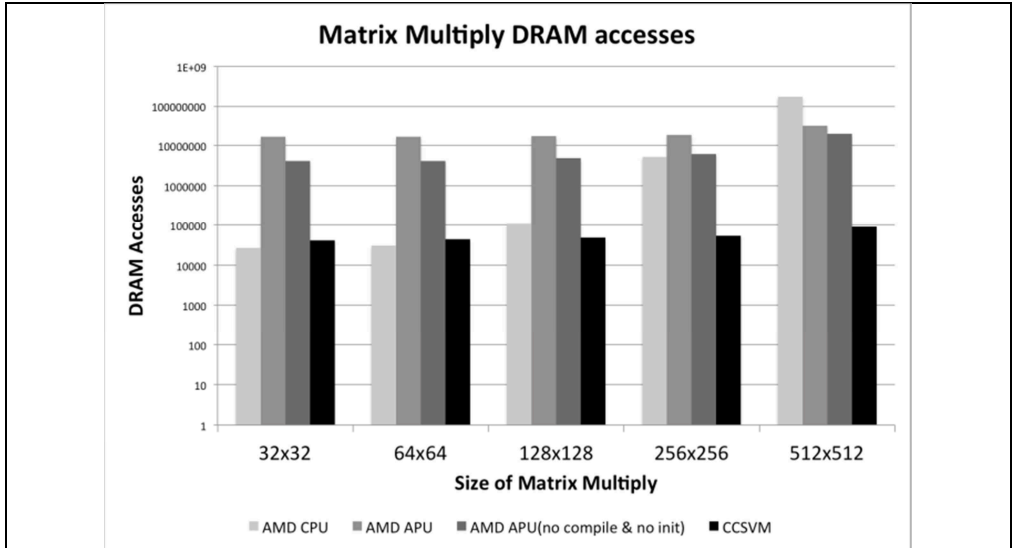

**Figure 9. DRAM Accesses for Matrix Multiply. CCSVM/xthreads avoids many off-chip accesses.**

Another potential challenge is that our HMC assumes write-back caches, even though current GPUs support write-through caches for reasons specific to graphics workloads. Future HMC designs will either have to adapt to write-through caches for CPUs, adapt write-back caches for GPUs, or provide a mechanism to change the write-back versus write-through policy based on the workload.

### 6.2 Scalability of Cache Coherence Protocol

We do not claim that the protocol in our CCSVM HMC is optimal; we did not tailor it in any way for a HMC. Future work will need to consider both tailored protocols as well as the scalability of such protocols to the vast number of cores (CPU and MTTOP) expected in future HMCs. Recent work indicates that scalable coherence is achievable [28][10], but it is not clear how to maintain scalability with a protocol that performs well for HMCs.

## 7 Related Work

In addition to the AMD APU that we have already discussed, there is a vast space of possible HMC memory system designs. In this section, we discuss a few other notable HMCs, plus we present other related work.

### 7.1 IBM PowerEN

IBM's PowerEN chip [3] is a heterogeneous multicore processor with Power-ISA CPU cores and several accelerator coprocessors. The coprocessors are fairly tightly coupled to the CPU cores and to each other, and all cores share virtual memory. To share virtual memory, the coprocessors must be able to perform address translation, and thus the PowerEN provides the coprocessors with an MMU for this purpose. Coprocessors are not full peers in the cache coherence protocol, but instead communicate with CPU cores via cache-coherent DMA.

### 7.2 Intel Prototypes: EXO and Pangaea

Intel's EXO prototype [40] consists of a multicore x86 CPU chip and a GPU chip, with shared virtual memory across the CPU and GPU cores. The paper focuses on the hardware prototype, which does not provide coherence, but later mentions a modeled but undescribed version of the design that provides coherence. Unlike typical homogeneous systems with CCSVM, the GPU cores do not have data caches.

Intel's Pangaea prototype [41] is a chip with CPU and GPU cores that share a virtual address space and share an L2 cache. As in EXO, the GPU cores do not have data caches, thus avoiding many coherence issues. Coherence between the GPU and the CPU cores' writeback L1 caches is enforced only at the request of software. Software can specify that a region of shared virtual memory is to be coherent, and only that region is kept coherent. Using software to manage coherent regions of memory is a key difference between Pangaea and today's CCVSM homogeneous chips which use hardware coherence.

### 7.3 Evaluations of CPU/GPU Chips

Researchers have recently compared the performance of workloads running on CPU/GPU chips and, in the process, have uncovered numerous performance issues. Gregg and Hazelwood [14] experimentally showed that the time to transfer data between CPU and GPU cores must be considered to avoid misleading conclusions; their results inspire work, such as ours, to reduce the overhead of communication between CPU and GPU cores. Similarly, Daga et al. [8] showed that the AMD APU's performance is far better than previous designs in which the CPU and GPU cores communicated over PCIe. Lee et al. [26] showed how certain performance evaluations have unfairly favored GPUs, with respect to CPUs.

## 8 Conclusions

We have demonstrated that the tight coupling of cores provided by CCSVM can potentially offer great benefits to an HMC. We do not claim that our CCVSM architecture, chip microarchitecture, or xthreads programming model are "optimal," assuming that optimality could even be defined. Rather, our CCSVM architecture with the xthreads programming model is a functional and promising starting point for future research into new features (e.g., programming language extensions) and optimizations for performance and efficiency (e.g., coherence protocols and consistency models tailored for heterogeneous chips, OS support for CCSVM on non-CPU cores, etc.).

## 9 Acknowledgments

This material is based on work supported by AMD and by the National Science Foundation under grant CCF-1216695.